\documentclass[sigconf]{acmart}

\renewcommand\footnotetextcopyrightpermission[1]{} %
\usepackage[final]{todonotes}   %
\usepackage{cleveref}
\usepackage{hyperref}
\usepackage{multirow}
\usepackage{booktabs} %
\usepackage{comgipp}
\acmDOI{10.475/123_4}

\acmISBN{123-4567-24-567/08/06}

\acmConference[JCDL'19]{ACM/IEEE Joint Conference on Digital Libraries}{Jun. 2019}{Urbana-Champaign, IL, USA}
\acmYear{2019}
\copyrightyear{2019}

\acmPrice{15.00}

\newcommand{\theReferenceText}{N.Meuschke, V.Stange, M.Schubotz, M.Kramer, and B.Gipp. Improving Academic Plagiarism Detection for STEM Documents by Analyzing Mathematical Content and Citations. In Proceedings ACM/IEEE Joint Conference on Digital Libraries (JCDL), 2019}
\begin{document}
\title[Improving PD for STEM Documents by Analyzing Mathematics and Citations]{Improving Academic Plagiarism Detection for STEM Documents by Analyzing Mathematical Content and Citations}
\author{Norman Meuschke\texorpdfstring{\textsuperscript{1,2}}{}, Vincent Stange\texorpdfstring{\textsuperscript{2}}{}, Moritz Schubotz\texorpdfstring{\textsuperscript{1}}{}, Michael Kramer\texorpdfstring{\textsuperscript{2}}{}, Bela Gipp\texorpdfstring{\textsuperscript{1,2}}{}}

\affiliation{
	\institution{
    \textsuperscript{1}University of Wuppertal, Germany (last@uni-wuppertal.de)
    }
}

\affiliation{
	\institution{
    \textsuperscript{2}University of Konstanz, Germany (first.last@uni-konstanz.de)
    }
}
\renewcommand{\shortauthors}{Meuschke, Stange, Schubotz, Kramer, Gipp}
\begin{abstract}
Identifying academic plagiarism is a pressing task for educational and research institutions, publishers, and funding agencies. Current plagiarism detection systems reliably find instances of copied and moderately reworded text. However, reliably detecting concealed plagiarism, such as strong paraphrases, translations, and the reuse of nontextual content and ideas is an open research problem. In this paper, we extend our prior research on analyzing mathematical content and academic citations. Both are promising approaches for improving the detection of concealed academic plagiarism primarily in Science, Technology, Engineering and Mathematics (STEM). We make the following contributions: i) We present a two-stage detection process that combines similarity assessments of mathematical content, academic citations, and text. ii) We introduce new similarity measures that consider the order of mathematical features and outperform the measures in our prior research. iii) We compare the effectiveness of the math-based, citation-based, and text-based detection approaches using confirmed cases of academic plagiarism. iv) We demonstrate that the combined analysis of math-based and citation-based content features allows identifying potentially suspicious cases in a collection of 102K STEM documents. Overall, we show that analyzing the similarity of mathematical content and academic citations is a striking supplement for conventional text-based detection approaches for academic literature in the STEM disciplines. The data and code of our study are openly available at
\href{https://purl.org/hybridpd}{\texttt{https://purl.org/hybridPD}}
\end{abstract}
\maketitle
\thispagestyle{myfirststyle}
\section{Introduction}\label{sec.intro}
Academic plagiarism (AP) is \textit{'the use of ideas, concepts, words, or structures without appropriately acknowledging the source to benefit in a setting where originality is expected'} \cite{fishman2009, Gipp2014a}. Forms of AP range from copying content (\textit{copy\&paste}) to reusing slightly modified content, e.g., interweaving text from multiple sources, to heavily concealing content reuse, e.g., by paraphrasing or translating text, and reusing data or ideas without proper attribution \cite{WeberWulff2014}.

The easily recognizable copy\&paste-type AP is more prevalent among students \cite{mccabe2005}, while concealed AP is more characteristic of researchers, who have strong incentives to avoid detection \cite{alzahrani2012}. Plagiarized research publications can have a severe negative impact by distorting the mechanisms for tracing and correcting research results and causing inefficient allocations of research funds \cite{Gipp2014a}. Therefore, detecting concealed AP in research publications is a pressing problem affecting many stakeholders, including research institutions, academic publishers, digital library providers, funding agencies, and of course other researchers \cite{Meuschke2018a}.

As we present in \Cref{sec.related_work}, many plagiarism detection (PD) approaches have been proposed that employ lexical, semantic, syntactical, or cross-lingual text analysis. These approaches reliably detect copied or moderately altered forms of AP; some approaches can also identify paraphrased and translated text. However, current approaches are computationally expensive and thus require a computationally less demanding selection of candidate documents before their application. The performance of methods that analyze textual features to retrieve candidate documents has reached a plateau. Therefore, the candidate retrieval step currently limits the effectiveness of PD approaches to detect concealed forms of AP.

Prior research (cf. \Cref{sec.related_work}) showed that approaches that analyze nontextual content features, such as academic citations, images, and mathematical content, complement the many text analysis approaches to improve the identification of concealed forms of AP. Nontextual content features in academic documents are a valuable source of semantic information that is largely independent of natural language text. Considering these sources of semantic information for similarity analysis raises the effort plagiarists must invest for obfuscating reused content \cite{Gipp2014a,Meuschke2014}.

We extend the research on analyzing nontextual content features for PD by devising a novel PD approach that combines the analysis of mathematical content with the analysis of academic citations. We structure the presentation of our contributions as follows. \Cref{sec.related_work} presents an overview of prior research on PD to show the benefit of analyzing nontextual content features for this purpose. \Cref{sec.mathpd} explains the conceptual design and technical realization of the math-based, citation-based and text-based PD approaches we investigate. \Cref{sec.eval} presents our test collection and describes the methodology of our evaluation. \Cref{sec.conf_cases} compares the performance of the three PD approaches using confirmed cases of AP. \Cref{sec.manual_investigation} presents the results of an exploratory study that investigates the ability of the novel PD approach to discover so far unknown cases of AP. \Cref{sec.concl} concludes the paper and presents future work.
\vspace{+0.45em}
\section{Related Work}\label{sec.related_work}
External plagiarism detection is an information retrieval task with the objective of comparing an input document to a large collection and retrieving all documents exhibiting similarities above a threshold \cite{Stein2007a}. External PD approaches typically employ a two-stage process consisting of candidate retrieval and detailed analysis \cite{Stein2007a,Meuschke2013}. In the candidate retrieval stage, the approaches employ computationally efficient retrieval methods to limit the collection to a set of documents that may have been the source for the content in the input document. In the detailed analysis stage, the systems perform computationally more demanding analysis steps to substantiate the suspicion and to align components in the input document and potential source documents that are similar \cite{Meuschke2013, alzahrani2012}.

Text retrieval research has yielded mature systems that reliably detect copied or moderately altered text in an input document and retrieve its source if the source is part of the system's reference collection. For the candidate retrieval stage, such systems typically employ character-gram or word-gram fingerprinting \cite{Oberreuter2011,Velasquez2016} or term-based vector space models %
\cite{K2015}. For the detailed analysis stage, such systems often perform exhaustive string comparisons %
\cite{Velasquez2016} or computationally more efficient text alignment %
\cite{Oberreuter2011}. Text alignment approaches typically use matching strings as seeds, which the procedures extend and then filter using heuristics \cite{Sanchez-Perez2014}.

To detect monolingual paraphrases, researchers have proposed approaches that analyze semantic and syntactic features, mostly during the detailed analysis stage of the PD process. Several researchers adapted semantic text analysis methods, such as Singular Value Decomposition \cite{Ceska2008}, Latent Semantic Analysis \cite{Soleman2014} and Explicit Semantic Analysis \cite{Meuschke2017} for the PD use case. Other PD approaches employ linguistic resources, such as WordNet\footnote{\url{https://wordnet.princeton.edu/}}, to analyze exactly matching and semantically related words %
\cite{Gupta2014}. Some works combine the analysis of word-based semantic similarity with an analysis of similarity in semantic arguments derived using Semantic Role Labeling %
\cite{Paul2015}. Other approaches employ part-of-speech tagging to also compare the syntactic structure of documents %
\cite{Gupta2014}.

To detect cross-lingual (CL) plagiarism, researchers have proposed approaches that leverage lexical similarities of languages, e.g., CL character $n$-gram matching, employ thesauri, parallel corpora, and machine translation followed by a mono-lingual analysis \cite{Barron-Cedeno2013}.

The extensive research on semantic and syntactic methods for monolingual and cross-lingual PD has yielded approaches that are highly effective for the detailed analysis stage. To our knowledge, \citeauthor{Gupta2014} reported the highest retrieval effectiveness for the detailed analysis of realistically obfuscated plagiarism \cite{Gupta2014}. The authors analyzed artificially created plagiarism in the Webis Text Reuse Corpus 2012 (Webis-TRC-2012) \cite{Potthast2012}, which is a standard corpus to evaluate PD systems as part of the PAN Workshop series\footnote{\url{http://pan.webis.de/}}. For manually paraphrased instances of simulated plagiarism, \citeauthor{Gupta2014} reported the precision $P=0.81$, recall $R=0.80$, and $F_1=0.80$.

While these results for the detailed analysis stage are promising, significant potential for improvement remains regarding the complete retrieval process. Semantic and syntactic analysis approaches like the one of \citeauthor{Gupta2014} are computationally expensive and thus require a prior candidate retrieval step. \citeauthor{Kong2013} achieved the best recall ($R=0.65$) for the candidate retrieval task in all four PAN workshops (2012-2015) that evaluated research contributions for this task \cite{Kong2013}. The organizers of the PAN workshop series noted that the workshops had reached a stable production phase, in which the submitted approaches no longer exhibited \textit{"[...] real innovations with respect to recall-oriented source retrieval."} \cite{Hagen2015}.

For cross-lingual PD, the candidate retrieval stage likewise seems to present an upper bound for the otherwise higher effectiveness of the analysis methods in the detailed analysis stage. \citeauthor{Ehsan2016} reported an approach that achieved an $F_1$-score of $0.87$ ($P=0.93$, $R=0.82$) for the detailed analysis of cross-lingual plagiarism in the Webis-TRC-2012 \cite{Ehsan2016}. For candidate retrieval, \citeauthor{Ehsan2016a} reported a maximum recall of $R=0.75$ using the same corpus \cite{Ehsan2016a}. These results suggest that the candidate retrieval step, which is necessary to enable applying semantic, syntactic and cross-lingual PD approaches, currently limits the effectiveness of the PD approaches.

PD approaches that analyze nontextual content features in academic documents are a promising complement to the variety of text analysis approaches both for the candidate retrieval and the detailed analysis stages of the PD process. In prior research, we showed that an analysis of in-text citation patterns in academic documents, i.e., identical citations occurring in proximity or in a similar order within two documents, can identify concealed forms of AP in real-world, large-scale collections \cite{Gipp2011,Gipp2011c,Gipp2014a,Gipp2014}. This approach is computationally efficient enough to be applied in the candidate retrieval stage \cite{Gipp2014a,Gipp2014,Meuschke2014}. \citeauthor{Pertile2016} confirmed the positive effect of combining citation and text analysis and devised a hybrid approach using machine learning \cite{Pertile2016}. The benefits of citation-based PD are twofold. First, citations encode semantic information that cannot easily be substituted, since leaving out citations to relevant prior work or citing sources that are not relevant to the topic would likely raise the suspicion of expert peer reviewers. Second, while citations are independent of natural language text, analyzing in-text citation patterns can indicate shared structural and semantic similarity among texts. Assessing this semantic and structural similarity using citation patterns requires significantly less computational effort than approaches for semantic and syntactic text analysis. 

We also showed that analyzing image similarity in academic documents, e.g., the similarity of figures and plots, improves the detection capabilities for concealed forms of AP \cite{Meuschke2018}.

In a recent short paper \cite{Meuschke2017b}, we extended the idea of citation-based PD. We proposed that mathematical expressions share many characteristics of academic citations and hence are promising nontextual content features to be considered when searching for concealed forms of AP. Similar to academic citations, mathematical expressions are essential components of academic documents in the Science, Technology, Engineering and Mathematics fields. Furthermore, mathematical expressions are independent of natural language text and contain rich semantic information. Additionally, some STEM disciplines, such as mathematics and physics, are known for their comparably sparse use of academic citations \cite{Moed85}. A citation-based analysis alone is, therefore, less likely to reveal potentially suspicious content similarity for these disciplines. 

Our piloting study \cite{Meuschke2017b} investigated measures to quantify the similarity of mathematical content features in the detailed analysis stage of the PD process. Given the infancy of the research area, we evaluated the suitability of comparing basic presentational features of mathematical expressions, i.e., elements of mathematical notation, such as identifiers, numbers, operators, and special symbols. Our goal was to identify the type of similar mathematics that we had observed in confirmed cases of AP, which we collected, e.g., by reviewing journal retractions. We embedded the retracted test documents together with their sources in the dataset of the NTCIR-11 Math task (105,120 arXiv documents) \cite{Aizawa14}. We performed pairwise comparisons of all documents in the dataset (detailed analysis approach) and evaluated similarity measures that consider identifiers, numbers, operators, and combinations thereof as the features. The best performing approach, a set-based comparison of the frequency of mathematical identifiers, retrieved eight of ten test cases at the top rank and achieved a mean reciprocal rank of $0.86$.
\newpage
The present paper extends our pilot study \cite{Meuschke2017b} by making four contributions: i) devising a candidate retrieval stage that analyzes mathematical expressions, citations, and textual features; ii) proposing new math-based similarity measures for the detailed analysis stage that consider the order of mathematical content features;\linebreak iii) comparing the effectiveness of the math-based, citation-based and text-based PD approaches using the confirmed cases of AP gathered for our pilot study; iv) analyzing math-based and citation-based content features to discover potentially suspicious cases of document similarity in a large-scale dataset of STEM documents.

\section{Detection Approach}\label{sec.mathpd}
\begin{figure*}[htb!]
\includegraphics[width=0.85\textwidth]{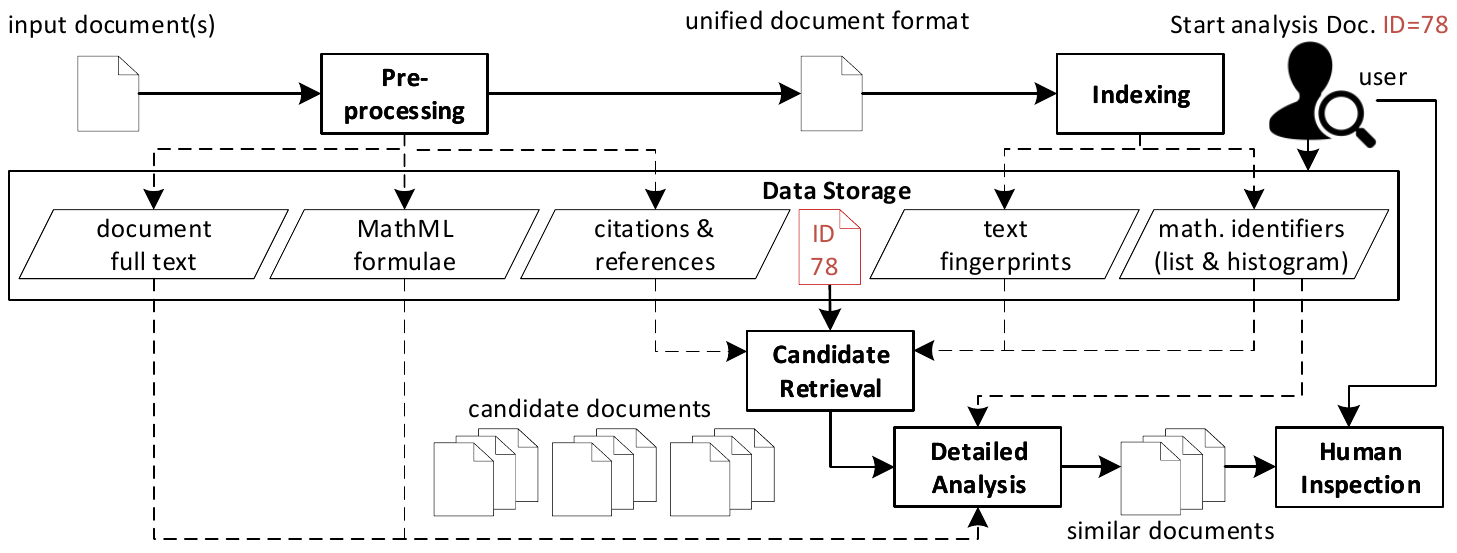}
\caption{Overview of the hybrid plagiarism detection approach.}\label{fig.detection_process}
\end{figure*}
\Cref{fig.detection_process} gives an overview of our system HyPlag \cite{Meuschke2018a} that implements the analysis of math-based, citation-based, and text-based content features. HyPlag allows the combination of the analysis approaches as part of a hybrid detection process that consists of five stages: preprocessing, indexing, candidate retrieval, detailed analysis, and human inspection. We describe the stages hereafter.

\subsection{Preprocessing}{\label{sec.mathpd.pre}}
HyPlag preprocesses input documents in two steps. In the first step, the system converts the documents to a unified XML-based document format used for the second preprocessing step. Our unified document format uses a subset of the TEI standard\footnote{\url{http://tei-c.org/}} defined by the information extraction tool GROBID\footnote{\url{https://github.com/kermitt2/grobid}} to represent in-text citations and bibliographic references. Additionally, the unified document format employs a subset of the Mathematical Markup Language (MathML)\footnote{\url{https://w3.org/Math/}} to represent mathematical formulae.

For this study, we processed documents in two formats: PDF (confirmed cases of AP) and LaTeX source code (NTCIR-11 MathIR Task dataset). We used GROBID to obtain bibliographic references from documents in both formats because the tool achieved excellent results for extracting header metadata, citations, and references \cite{Tkaczyk2015,Bast2017}. Since GROBID cannot recognize mathematical formulae, we semi-automatically invoked InftyReader\footnote{\url{http://www.inftyreader.org/}} to convert the PDFs for confirmed cases of AP to an intermediate LaTeX format. InftyReader is currently the most commonly used OCR-based recognition system for mathematical content \cite{Iwatsuki2017}. While the tool typically achieves a recall of at least $0.90$, the precision can be as low as approx. $0.15$ in some scenarios \cite{Iwatsuki2017}. To prevent bias from recognition errors, we manually checked and corrected the LaTeX output of InftyReader.

We employed LaTeXML\footnote{\url{https://dlmf.nist.gov/LaTeXML}} to convert LaTeX documents, i.e., the documents in the NTCIR-11 dataset and the PDF that InftyReader converted to LaTeX, to the unified document format. The LaTeXML library offers mathematical content conversion from LaTeX source code to a MathML representation. To enable conversion to our unified document format, we contributed an XSL style sheet that transforms LaTeXML's native output to TEI. The new conversion option has been included in the LaTexML distribution\footnote{\url{https://github.com/brucemiller/LaTeXML/blob/master/lib/LaTeXML/resources/XSLT/LaTeXML-tei.xsl}}. 

To recognize in-text citations, LaTeXML requires the use of LaTeX tags, such as \verb|\cite{}|. Many documents in the dataset do not contain such markup but state in-text citations as plain-text. In such cases, our preprocessing pipeline did not recognize the in-text citations. Additionally, many documents do not use in-text citations at all but only reference items in the bibliography. Due to both errors, the number of unique in-text citations for 68,743 documents (67\% of the dataset) is smaller than the number of references.

In the second preprocessing step, HyPlag splits the unified document format into separate data structures holding plain text, mathematical formulae, in-text citations, and bibliographic references. To extract plain text, the system removes all XML structures, images, formulae, and formatting instructions. Formulae in Content MathML are extracted as they are. In-text citations are linked to the corresponding reference entries in the bibliography; reference entries are split into author, title, and venue fields.

\subsection{Indexing}
In the indexing stage, HyPlag stores into an Elasticsearch\footnote{\url{https://www.elastic.co/products/elasticsearch}} index the following data extracted from the preprocessed documents: \textit{Document metadata}: title, authors, publication date and filename. 

\textit{Mathematical features}: the sequence of all mathematical identifiers in the order of their occurrence in the document, and the unordered histogram of the occurrence frequencies of identifiers, i.e., how often an identifier occurs in the document. We focused on analyzing identifiers since they achieved the best retrieval effectiveness in our pilot study \cite{Meuschke2017b}. We used the MathML \verb|<formula>| element to extract formulae. Additionally, we used \verb|<ci>| elements in MathML formulae to extract identifiers in a formula.

\textit{Citation features}: bibliographic references and in-text citations. The system consolidates the data about referenced documents by comparing the title and author names extracted from reference strings to the data of previously indexed documents while accounting for minor spelling variations utilizing the Levenshtein distance.

\textit{Textual features}: full text and text fingerprints formed by chunking the document into word 3-grams and applying probabilistic chunk selection (average chunk retention rate $1/16$). To realize the text fingerprinting approach, we adapted the Sherlock\footnote{The tool's website went offline recently. The source code and documentation are still available via the Web archive: \url{https://web.archive.org/web/20180219024142/http://web.it.usyd.edu.au/~scilect/sherlock/}} tool.

The indexing process is identical for all documents, i.e., documents that ought to be analyzed need to be indexed first.

\subsection{Candidate Retrieval} \label{sec.candidate_retrieval}
In the candidate retrieval stage, the system queries the index using mathematical identifiers, in-text citations, and text fingerprints extracted from an input document to retrieve a set of candidate documents for the subsequent detailed analysis.

To retrieve candidate documents, we employed "Lucene's practical scoring function" implemented in the Elasticsearch server as a computationally efficient, well-established heuristic. The scoring function combines a tf/idf weighted vector space model with a Boolean retrieval approach\footnote{documentation of the scoring function: \url{https://www.elastic.co/guide/en/elasticsearch/guide/1.x/practical-scoring-function.html}}. 

We performed three queries, each retrieving the 100 documents with the highest relevance scores. For the citation-based and text-based retrieval of candidate documents, in-text citations and, respectively, text fingerprints of the input document represented the terms of the query. Analogously, indexed documents were represented by their sets of in-text citations and text fingerprints. We used the default parameters of Lucene's scoring function. 

For the math-based retrieval of candidate documents, the set of mathematical identifiers occurring in a document was the query. The indexed documents were represented by the sequence of mathematical identifiers in a document, i.e., identifiers can occur more than once. However, using Lucene's default parameters for the relevance scoring yielded unsatisfactory results in the case of mathematical features. This finding is in line with research by \citeauthor{Sojka2011} \cite{Sojka2011}. Similar to \citeauthor{Sojka2011}, we found that query terms, i.e., mathematical identifiers, should be given additional weight for multiple occurrences. Therefore, we set the boost value $\operatorname{boost}(t)$ for the term $t$ in the query, i.e., individual identifiers, to the number of occurrences of the term (identifier) in the query document. 

Since we sought to investigate the effectiveness of the math-based, the citation-based, and the text-based PD approach independently of each other, we did not consolidate the three sets of 100 candidate documents each retrieved by the three queries.

\subsection{Detailed Analysis} \label{sec.detailed_analysis}
In the detailed analysis stage, HyPlag compares the input document(s) to all documents in each of the three sets of 100 candidate documents retrieved in the previous stage. For each document comparison, HyPlag computes the following similarity measures.

\subsubsection{Math-based similarity measures} We only computed math-based similarity scores for document pairs that share 20 or more identifiers to prevent high similarity scores resulting from a few shared identifiers, such as the occurrence of $x$ and $y$. For documents that meet this threshold, we computed three similarity measures. 

First, we computed the similarity of \textit{frequency histograms of mathematical identifiers (Histo)}, which performed best in our pilot study \cite{Meuschke2017b}. The measure quantifies the similarity of two documents $d$ and $d'$ as the difference in the relative occurrence frequencies of identifiers $f_i$ in $d$ and $d'$ according to \Cref{eq.dist}.
\begin{equation} \label{eq.dist}
s(d,d')=1-\frac{\sum_{i\in I} \left|f_{i,d} - f_{i,d'}\right|}{\sum_{i\in I} \max(f_{i,d},f_{i,d'})}.
\end{equation}
The Histo score reflects the global overlap of identifiers in two documents. The measure is most suitable for documents with comparable numbers of identifiers. Typically, this requirement is not met if the two documents strongly differ in length.

In addition to the set-based, order-agnostic Histo measure proposed in our pilot study \cite{Meuschke2017b}, we devised two new similarity measures that consider the order of mathematical identifiers. We did not evaluate the influence of the sequential similarity of features in our pilot study. The two new measures consider the \textit{Longest Common Subsequence of Identifiers (LCIS)} and the set of \textit{Greedy Identifier Tiles (GIT)} for score computation. The longest common subsequence and greedy tiling algorithms are well-established approaches to identify sequential patterns and have been applied successfully for the text-based \cite{Jayapal2012} and citation-based \cite{Gipp2011c} detection of academic plagiarism, as well as for source code plagiarism detection \cite{Prechelt2002}.

The longest common subsequence of features, e.g., characters, citations, or mathematical identifiers, is the maximum number of features that match in both documents in the same order but not necessarily in a contiguous block. Like Histo, the LCIS measure quantifies the global similarity of documents. We compute the similarity score $s_\mathrm{LCIS}(d,d')=\left|L(d,d')\right|I_{d}^{-1}$ that represents the number of identifiers in the query document $I_{d}$ that are part of the longest common identifier sequence whose length is given by $L$.

Greedy tiles are the set of all individually longest blocks of shared features in identical order that cannot be extended to the left or right without encountering a non-matching feature \cite{Wise93}. Greedy tiles are well-suited to identify confined regions with high similarity \cite{Gipp2011c}. We computed the similarity of two documents using the GIT approach as
$s_\mathrm{GIT}(d,d')=\left|T_l\right|I_{d}^{-1}$, where $T_l$ is the set of tiles with a length greater or equal to 5 matching identifiers and $I_{d}$ is the total number of identifiers in the query document. In other terms, the score quantifies the number of identifiers in the query document that are part of identifier tiles with a minimum length of five. 

\subsubsection{Text-based similarity assessment}
For the detailed text-based analysis, we used the \textit{Encoplot algorithm (Enco)} developed by \citeauthor{Grozea09} \cite{Grozea09}. Encoplot is an efficient character 16-gram comparison that achieves $O(n)$ time-complexity by ignoring repeated matches. The similarity score is the ratio of shared character 16-grams to all 16-grams of the shorter document.

\subsubsection{Citation-based similarity assessment} We used three approaches that proved effective in our prior research\cite{Gipp2014a,Gipp2014,Gipp2011,Gipp2011c,Meuschke2014}. 

\textit{Bibliographic Coupling (BC)}, quantifies the fraction of shared bibliographic references. The similarity score is calculated as\linebreak
$s(d,d')=\left|R_{d}\cap R_{d'}\right|\left(R_{d}\cup R_{d'}\right)^{-1}$, where $R_{d}$ and $R_{d'}$ are the sets of references in the query and the comparison document. Like the Histo measure for mathematical features, BC is an order-agnostic measure that quantifies the global citation-based similarity of documents. It achieves high scores if documents with similar numbers of references share a significant fraction of those references.

The \textit{Longest Common Citation Sequence (LCCS)} and \textit{Greedy Citation Tiling (GCT)} measures follow the same idea as LCIS and GIT but consider in-text citations instead of mathematical identifiers. The similarity scores for the LCCS and GCT approaches are calculated analogously to the scores for LCIS and GIT, with the exception that the minimum length for citation tiles is two matching citations opposed to five matching identifiers for GIT. We only computed the three citation-based similarity measures if both documents we compared contained at least three bibliographic references.

\subsection{Human Inspection}
We used HyPlag's web-based frontend to visualize content similarity for inspection. For details on the visualizations, see \cite{Meuschke2018a}. 

\section{Evaluation}\label{sec.eval}
To ensure the reproducibility of our research, the data and code of our study are available at \href{https://purl.org/hybridpd}{\texttt{https://purl.org/hybridPD}}.
\subsection{Dataset}
To achieve comparability to our prior research, we reused the dataset of our previous experiments \cite{Meuschke2017b}. The dataset consists of ten cases that we selected after manually reviewing 44 research publications in STEM disciplines that have been officially retracted for plagiarism and involve mathematical content. We restricted the dataset to ten cases for three reasons. First, we chose cases from research fields within our area of expertise to enable us to assess the severity of identified similarities. Second, we selected cases that are most representative of the types of mathematical similarity we observed. Third, the effort required for checking the output of InftyReader and correcting incorrectly recognized mathematical expressions prevented us from converting more cases. 

We chose using real cases of AP over creating artificial test cases, although gathering and converting real cases is time-consuming and thus resulted in a smaller dataset. The reason is that we see the ability to identify real cases of AP committed by researchers who are experts in their fields and have a strong incentive to avoid detection as the ultimate performance test for any PD approach. Therefore, we see real cases of AP as best suited to devise and evaluate the novel hybrid detection approach. 

We embedded the ten retracted documents and the ten source documents for our test cases in the topically related NTCIR-11 MathIR Task dataset. The NTCIR dataset, which is available for research purposes \cite{Aizawa14}, consists of 105,120 scientific papers in LaTeX format from computer science, mathematics, physics, and statistics that were published in the arXiv preprint repository\footnote{\url{http://www.arxiv.org}}. 

Using our preprocessing pipeline (cf. \cref{sec.mathpd.pre}), we converted all LaTeX source files of the NTCIR dataset and of our test cases to HyPlag's unified document format. We excluded 2,616 documents, for which LaTeXML or our TEI parser encountered critical processing errors.
Approximately one-third of the remaining documents did not contain markup for authors and title. To achieve the best possible data quality, we used the arXiv API\footnote{\url{https://arxiv.org/help/api/}} to obtain author and title information for all documents instead of extracting the information from the LaTeX source files. %
For 6,770 documents we were unable to extract bibliographic references due to missing markup.
Since the arXiv API does not offer the data of bibliographic reference, we indexed these documents without reference data.

\Cref{tab.features} shows the number of content features we obtained for the final dataset of 102,524 documents. The numbers confirm that this collection of STEM documents contains a significantly higher number of mathematical formulae (52M) than academic citations (3M). Therefore, analyzing both mathematical formulae and citations is more promising in these disciplines than analyzing citations alone. The formulae contain more than 156M identifiers, which the system grouped into 4,063,354 identifier histogram entries. On average, documents contained 70 different mathematical identifiers.
\begin{table}
\caption{Overview of content features in our dataset.}
\label{tab.features}
\begin{tabular}{@{}lrr@{}}
\toprule
\textbf{Features} & \textbf{Total} & \textbf{Avg. per doc.} \\ \midrule
references     & 2,201,094      & 21                   \\
unique references & 1,445,059       & --                       \\
citations      & 3,068,865      & 30                   \\
text fingerprints  & 26,539,276     & 256                  \\
formulae       & 52,271,908     & 504                  \\
math. identifiers   & 156,706,600      & 1513                   \\
\end{tabular}
\vspace{-.5cm}
\end{table}
\subsection{Investigations} \label{sec.investigations}
To evaluate the effectiveness of the PD approaches, we performed two conceptually different investigations. The first investigation reflects the typical scenario in external PD, i.e., checking an input document for similarity to documents in a collection. We submitted the retracted paper for each test case to our system HyPlag. For each query document, the system used the math-based, citation-based, and text-based retrieval heuristics (cf. \Cref{sec.candidate_retrieval}) to retrieve three sets of 100 candidate documents each. In the subsequent detailed analysis stage, each query document was compared to all the candidate documents in the three sets without consolidating the sets. \Cref{sec.conf_cases} presents the results of this investigation. 

The second investigation assesses the effectiveness of combining the math-based and citation-based similarity measures to discover so far unknown cases of potentially suspicious document similarity. We submitted each of the $N=102,524$ documents in our dataset to HyPlag. We retrieved the three sets of candidate documents by applying the math-based, citation-based, and text-based retrieval heuristics for all $N$ documents. Opposed to the evaluation of confirmed plagiarism cases, we formed the union of the sets to enable the exploration of approaches that combine measures. In the detailed analysis stage, we compared each of the $N$ documents in the dataset to its consolidated set of candidate documents $C$. We manually examined the retrieved documents with the highest similarity scores. \Cref{sec.manual_investigation} presents the results of this investigation.

\section{Results Confirmed Cases of AP}\label{sec.conf_cases}
To not establish a link between a paper on academic plagiarism \mbox{detection} and legitimate research papers, i.e., the source documents of our test cases and unsuspicious documents retrieved in our experiments, we do not cite the documents we discuss hereafter. However, all documents are accessible via our data and code repository.
\subsection{Candidate Retrieval}
\Cref{tab.candidateRet} shows the effectiveness of the candidate retrieval approaches. Plus signs ($+$) in the table indicate that HyPlag retrieved the source document among the 100 candidate documents when the retracted document for each of the ten test cases (C1 \ldots C10) was the query. Minus signs ($-$) indicate that an analysis approach did not retrieve the source document among the candidate documents. The rightmost column in \Cref{tab.candidateRet} shows the recall of the approaches.

\begin{table}[htb!]
\centering
\caption{Recall for candidate retrieval stage.}
\label{tab.candidateRet}
\begin{tabular}{@{}l|p{0.25cm}p{0.25cm}p{0.25cm}p{0.25cm}p{0.25cm}p{0.25cm}p{0.25cm}p{0.25cm}p{0.25cm}p{0.5cm}|r@{}}
\toprule
              & \textbf{C1} & \textbf{C2} & \textbf{C3} & \textbf{C4} & \textbf{C5} & \textbf{C6} & \textbf{C7} & \textbf{C8} & \textbf{C9} &
              \textbf{C10} & $\textbf{R}$   \\
              \midrule
\textbf{Math} & +           & +           & +           & --          & --           & --          & +           & +           & +           & +            & {0.7} \\
\textbf{Cit.} & +           & +           & --          & +           & +           & +          & +           & +           & +           & +           & {0.9} \\
\textbf{Text} & +           & +           & +           & +           & +           & +          & --          & +           & +           & +            & {0.9} \\ %
\end{tabular}
\end{table}

Both the citation-based and the text-based approaches achieved a recall of $0.9$; the math-based approach achieved a recall of $0.7$. Notably, the three approaches failed to retrieve the source document among the candidates for distinct sets of test cases. Combining the three sets of candidate documents would result in a perfect recall.

\subsection{Detailed Analysis}
To quantify the effectiveness of the similarity measures employed in the detailed analysis stage, we performed a score-based assessment and a rank-based assessment.
\subsubsection{Score-based assessment} This assessment determines which scores are significant, i.e., potentially suspicious, for our similarity measures and dataset. To our knowledge, no study (including our pilot study \cite{Meuschke2017b}) has quantified the mathematical similarity that can be expected by chance to derive a significance threshold.

To establish significance thresholds for the scores of all similarity measures, we analyzed a random sample of 1 million document pairs as follows. We randomly picked two documents from the dataset. If the chosen documents had (a) common author(s) or if one of the documents cited the other, we discarded the pair. We continued the process until reaching the number of 1M document pairs. The selection criteria ought to eliminate document pairs that exhibit high content similarity for likely legitimate reasons, i.e., reusing own work and referring to the work of others with due attribution. Our goal was to estimate an upper bound for similarity scores that likely result from random feature matches. To do so, we manually assessed the topical relatedness of the top-ranked document pairs within the random sample of 1M documents for each similarity measure. We picked as the significance threshold for a similarity measure the rank of the first document pair for which we could not identify a topical relatedness. \Cref{tab.score_thresholds} shows the significance scores we derived using this procedure. 

\begin{table}[t!]
\caption{Significance thresholds for similarity measures.} \label{tab.score_thresholds}
\begin{tabular}{@{}llllllll@{}}
\toprule
             & \textbf{Histo} & \textbf{LCIS} & \textbf{GIT} & \textbf{BC} & \textbf{LCCS} & \textbf{GCT} & \textbf{Enco} \\ \midrule
\textbf{$s$} & $\geq.56$       & $\geq.76$      & $\geq.15$     & $\geq.13$    & $\geq.22$      & $\geq.10$     & $\geq.06$ \\     
\end{tabular}
\vspace{-.3cm}
\end{table}

\begin{figure} [b!]
\vspace{-0.2cm}
\includegraphics[width=\columnwidth]{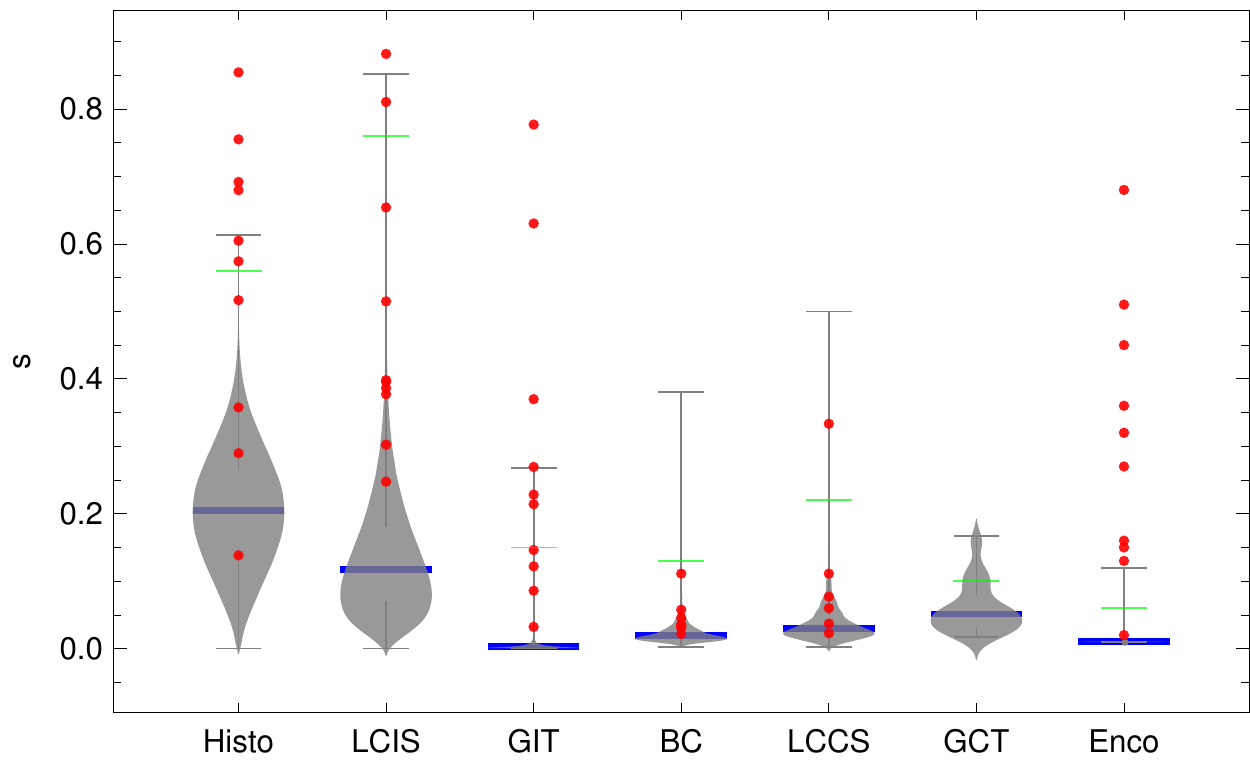} 
\caption{Similarity scores in 1M random document pairs.} \label{fig.dist_scores}
\label{fg.stat}
\end{figure}

\Cref{fig.dist_scores} shows the distribution of the similarity scores $s$ (vertical axis) computed using each similarity measure for the random sample of 1M documents. Large horizontal bars shaded in blue indicate the median score; small horizontal bars shaded in grey mark the minimum and maximum scores; small horizontal bars shaded in green indicate the significance thresholds for each measure (cf. \Cref{tab.score_thresholds}). The grey shapes in the chart show the smoothed probability density functions of the score frequencies, which were generated by applying a kernel-based density estimation. Red dots in the plot indicate the similarity scores of test cases for which the respective measure was applied, i.e., if the document pairs contained enough features to compute a score (cf. \Cref{sec.detailed_analysis}).

\begin{table*}[t!]
\centering
\caption{Retrieval effectiveness of detection approaches for confirmed cases of plagiarism.}
\label{tab.rankingPerformance}
\begin{tabular}{@{}lccccccccccccccccc@{}}
\toprule
& \multicolumn{6}{c}{\textbf{Math}}& \multicolumn{9}{c}{\textbf{Citation}} & \multicolumn{2}{c}{\textbf{Text}}\\ \cmidrule(l){2-7} \cmidrule(l){8-16} \cmidrule(l){17-18} 
\textbf{Case}       & \multicolumn{2}{c}{\textbf{Histo}}       & \multicolumn{2}{c}{\textbf{LCIS}}        & \multicolumn{2}{c}{\textbf{GIT}}         & \multicolumn{3}{c}{\textbf{BC}}                     & \multicolumn{3}{c}{\textbf{LCCS}}                   & \multicolumn{3}{c}{\textbf{GCT}}                    & \multicolumn{2}{c}{\textbf{Enco}}        \\ \midrule
\multicolumn{1}{c}{} & \textit{r}          & \multicolumn{1}{c|}{\textit{s}}         & \textit{r}          & \multicolumn{1}{c|}{\textit{s}}         & \textit{r}          & \multicolumn{1}{c|}{\textit{s}}         & \textit{r} & \textit{s} & \multicolumn{1}{c|}{\textit{s*}} & \textit{r} & \textit{s} & \multicolumn{1}{c|}{\textit{s*}} & \textit{r} & \textit{s} & \multicolumn{1}{c|}{\textit{s*}} & \textit{r}          & \textit{s}         \\
C1 & 1 & \multicolumn{1}{c|}{\underline{.68}} & 1 & \multicolumn{1}{c|}{.40} & 1 & \multicolumn{1}{c|}{\underline{.21}} & 1 & .06 & \multicolumn{1}{c|}{\underline{.15}} & 1 & .06 & \multicolumn{1}{c|}{.10} & - & - & \multicolumn{1}{c|}{.04} & 1 & \underline{.13} \\
C2 & 1 & \multicolumn{1}{c|}{\underline{.60}} & 1 & \multicolumn{1}{c|}{.39} & 1 & \multicolumn{1}{c|}{.12} & 10' & .05 & \multicolumn{1}{c|}{\underline{.28}} & 1 & \underline{.33} & \multicolumn{1}{c|}{\underline{.42}} & - & - & \multicolumn{1}{c|}{-} & 1 & \underline{.16} \\
C3 & 3 & \multicolumn{1}{c|}{.29} & 1 & \multicolumn{1}{c|}{\underline{.88}} & 1 & \multicolumn{1}{c|}{\underline{.78}} & - & - & \multicolumn{1}{c|}{-} & - & - & \multicolumn{1}{c|}{-} & - & - & \multicolumn{1}{c|}{-} & 1 & \underline{.36} \\
C4 & (1) & \multicolumn{1}{c|}{(.36)} & (99) & \multicolumn{1}{c|}{(.37)} & (3) & \multicolumn{1}{c|}{(.03)} & - & - & \multicolumn{1}{c|}{\underline{.35}} & - & - & \multicolumn{1}{c|}{\underline{.44}} & - & - & \multicolumn{1}{c|}{\underline{.25}} & 1 & \underline{.15} \\
C5 & (1) & \multicolumn{1}{c|}{(\underline{.57})} & (86) & \multicolumn{1}{c|}{(.30)} & (1) & \multicolumn{1}{c|}{(\underline{.23})} & 5 & .02 & \multicolumn{1}{c|}{\underline{.18}} & 7' & .02 & \multicolumn{1}{c|}{\underline{.23}} & - & - & \multicolumn{1}{c|}{.05} & 1 & \underline{.45} \\
C6 & (19) & \multicolumn{1}{c|}{(.14)} & (98) & \multicolumn{1}{c|}{(.40)} & (1) & \multicolumn{1}{c|}{(\underline{.15})} & 2 & .04 & \multicolumn{1}{c|}{\underline{.32}} & 1 & .11 & \multicolumn{1}{c|}{\underline{.44}} & - & - & \multicolumn{1}{c|}{\underline{.22}} & 1 & \underline{.27} \\
C7 & 2 & \multicolumn{1}{c|}{.52} & 98 & \multicolumn{1}{c|}{.25} & 1 & \multicolumn{1}{c|}{.09} & - & - & \multicolumn{1}{c|}{.04} & - & - & \multicolumn{1}{c|}{.05} & - & - & \multicolumn{1}{c|}{-} & (4) & (.02) \\
C8 & 1 & \multicolumn{1}{c|}{\underline{.76}} & 1 & \multicolumn{1}{c|}{.65} & 1 & \multicolumn{1}{c|}{\underline{.37}} & 1 & .11 & \multicolumn{1}{c|}{\underline{.37}} & - & - & \multicolumn{1}{c|}{\underline{.25}} & - & - & \multicolumn{1}{c|}{-} & 1 & \underline{.32} \\
C9 & 1 & \multicolumn{1}{c|}{\underline{.69}} & 1 & \multicolumn{1}{c|}{.51} & 1 & \multicolumn{1}{c|}{\underline{.27}} & 1 & .03 & \multicolumn{1}{c|}{\underline{.26}} & 1 & .08 & \multicolumn{1}{c|}{\underline{.39}} & - & - & \multicolumn{1}{c|}{-} & 1 & \underline{.68} \\
C10 & 1 & \multicolumn{1}{c|}{\underline{.85}} & 1 & \multicolumn{1}{c|}{\underline{.81}} & 1 & \multicolumn{1}{c|}{\underline{.63}} & 1 & .03 & \multicolumn{1}{c|}{.03} & 1 & .04 & \multicolumn{1}{c|}{.04} & - & - & \multicolumn{1}{c|}{-} & 1 & \underline{.51} \\ \midrule
\multirow{2}{*}{\textbf{MRR}} & .58   & & .60   & & .79   & & .48   & & & .60   & & & .00    & & & .90 & \\
                             & (.79) & & (.60) & & (.93) & & (.48) & & & (.60) & & & (.00)  & & & (.93) \\
\bottomrule
\end{tabular}
\end{table*}

As shown in \Cref{fig.dist_scores}, the probability density function (PrDF) of Histo is symmetrical while the PrDF for any other measure is negatively skewed, i.e., exhibits the highest frequencies at lower scores. The stronger the PrDF of scores is negatively skewed, the more selective the measure is. For the math-based similarity measures (Histo, LCIS, GIT), considering the order of identifiers strongly increases the selectivity of the measures. The PrDF for the order-agnostic Histo measure is symmetrical. The PrDF of scores for the LCIS measure, which leniently considers the order of identifiers in the entire document, is slightly skewed towards lower values, while the PrDF for the GIT measure, which focuses on considering identifier order, is strongly skewed towards lower values.

Given our prior research on citation-based similarity \cite{Gipp2011c, Gipp2014a, Gipp2014, Gipp15b}, we expected similar characteristics for the citation-based measures. However, as shown in \Cref{fig.dist_scores}, the order-agnostic BC measure is more selective than the order-considering LCCS measure in this case. The reason is the errors in citation extraction (cf. \Cref{sec.mathpd.pre}). The mismatch of references and in-text citations causes that the LCCS and GCT measures can only consider a fraction of the citations in the dataset. This fraction is smaller than the fraction of extracted references, which the BC measure uses. Therefore, the BC measure is more selective than the LCCS measure for this dataset, since overlaps of the comparably sparse in-text citations increased the LCIS score more than overlaps in references increased the BC score. Unrecognized in-text citations also cause the GCT measure to be overly selective for this dataset. Due to a shortage of data points, the PrDF for scores of the GCT measure shows interpolation artifacts, i.e., the PrDF is not monotonically decreasing for larger scores. HyPlag could identify in-text citation tiles above the exclusion thresholds (cf. \Cref{sec.detailed_analysis}) for only 41 document pairs in our sample of 1M documents and for none of our test cases. 

The PrDF of the Encoplot scores shows that the text-based measure is highly selective. Nine of the test cases have scores above the significance threshold, i.e., most verified cases of AP have a significant textual overlap with the respective source document. This characteristic is common for confirmed cases of AP \cite{Potthast2010a}. Identifying literal text overlap is easier for reviewers and better supported by productive PD systems than identifying concealed content similarity. Therefore, documents with (near) copied text are more likely to be discovered and hence likely overrepresented in our dataset.
\subsubsection{Combined rank-based \& score-based assessment.} In addition to assessing the significance of the similarity scores, we also examined the ranks $r$ at which the similarity measures retrieved the source document for each of the test cases. To indicate the average ranking performance of the measures, we computed the \textit{Mean Reciprocal Rank (MRR)}. In the case of tied ranks, we considered the mean rank, i.e., the pessimistically rounded average of the number of document pairs that share the same rank. The best possible score of 1 is assigned if a similarity measure exclusively retrieves the source document at rank 1 for each test case. 

\Cref{tab.rankingPerformance} shows the results of both the rank-based and the score-based assessment. For each of the test cases (C1\ldots C10) the table lists the rank $r$ at which HyPlag retrieved the source document and which score $s$ the similarity measure assigned. We mark the mean rank, which we list in the case of tied ranks, with an apostrophe, e.g., 7'. Scores above the significance threshold of a measure (see \Cref{tab.score_thresholds}) are underlined. To gauge the performance of the similarity measures specifically for the detailed analysis stage, we also state the ranks and similarity scores for the cases not retrieved in the candidate retrieval stage. We mark such entries with parentheses, e.g., (0.15). To compute the ranks and scores for these documents, we performed a comparison of the query document to all documents in the dataset. Minus characters ($-$) indicate that HyPlag computed no similarity score due to the exclusion criteria of the measure. Because of the incomplete and error-prone extraction of bibliographic data, we state a separate score $s^*$ for the citation-based measures. The score indicates the true citation-based similarity of the test cases. To compute $s^*$, we manually corrected erroneous data for in-text citations and references before applying the similarity measures.

The text-based approach consisting of word 3-gram fingerprinting (Sherlock) for the candidate retrieval stage and efficient string matching (Encoplot) for the detailed analysis stage achieved the best individual result. The approach retrieved nine of the ten test cases at the top rank. Only test case C7 exhibits a textual similarity that is too low to retrieve the source document in the candidate retrieval stage and mark the document as suspicious in the detailed analysis stage. The Encoplot scores for 6 of the 10 test cases exceed 0.25, hence are clearly suspicious. For the cases C1, C2, and C4, the Encoplot scores exceed our significance threshold of 0.06, yet are lower than 0.20. Anecdotal evidence\footnote{\url{www.researchgate.net/post/What_percentage_of_plagiarism_is_generally_treated_as_acceptable}} suggests that 10\% - 20\% of text overlap is not immediately suspicious but often tolerated by journal reviewers and editors. The practices regarding acceptable text overlap vary between research fields and even between venues. Whether a productive text-based PD system would flag C1, C2, and C4 as suspicious is thus unclear. The retraction note of C1 names the unattributed reuse of a mathematical analysis, not the textual overlap with the source, as the reason for the retraction. The scores for Histo (0.68) and Git (0.21), which both exceed the significance thresholds, reflect this similarity in mathematical content.

The math-based similarity measures achieved the second-best result when considering both the candidate retrieval and detailed analysis stages. GIT, which we devised as a new similarity measure for this study, performed particularly well, retrieving seven cases at the top rank. When only considering the detailed analysis stage, GIT achieved the same effectiveness as the text-based analysis (9~test~cases retrieved at rank one, MRR=0.93). To enable this result for the detailed analysis stage, the candidate retrieval procedure could simply combine the results of the math-based, citation-based and text-based approaches as discussed in \Cref{sec.candidate_retrieval}.

GIT outperformed the Histo measure, which achieved the best results in our pilot study \cite{Meuschke2017b}. In this prior study, Histo achieved an MRR score of $0.86$. Our current implementation exhibits a slightly lower MRR of 0.79. We attribute the difference to using a different conversion and data extraction process. The good performance of GIT suggests that the pattern of reusing (nearly) identical content in confined parts of a document known as "shake\&paste" or "patchwriting" \cite{WeberWulff2014} also applies to mathematical content.

For our test cases, LCIS achieved no significant improvement over the set-based Histo measure. Both LCIS and Histo achieved good results for test cases that share a large fraction of their mathematical content. For such documents, the amount of shared math sufficed to retrieve the documents using the Histo approach. That the large overlap in mathematical content also yielded long identifier subsequences did not significantly improve the similarity score.

The citation-based measures achieved the lowest overall performance, largely due to the deficiencies of the extracted data. Despite the sub-optimal data, the LCCS measure retrieved 5 cases at rank one achieving an MRR score of 0.60. The similarity scores $s^*$, which assume the bibliographic data in the documents would have been extracted and matched correctly, give a better indication of the potential effectiveness of the citation-based measures. Notably, LCCS would yield scores of approx. two times the significance threshold of 0.22 and hence strongly suspicious for C2, C4, C6, and C9. Given that C2 and C4 exhibit a textual overlap that is significant but not strongly suspicious (0.16 and 0.15), the high LCCS score could provide an indicator for suspicious similarity. 

For all cases except C7, which none of the measures flags as suspicious, at least one math-based or citation-based measure yields a similarity score above the individual significance thresholds. For Case C7, the Histo score has the smallest difference to the measure's significance threshold making Histo the most likely measure to retrieve the case despite the comparably low score. 

In summary, the evaluation using confirmed cases of AP showed that the combined analysis of math-based and citation-based similarity identified all cases that also a text-based analysis flagged as strongly suspicious. Moreover, the two nontextual detection approaches provide valuable indicators for suspicious document similarity for cases with a comparably low textual similarity.
\section{Exploratory Study Results}\label{sec.manual_investigation}
In this section, we describe our findings from manually investigating the top-ranked documents that HyPlag retrieved when applying math-based and citation-based content features to compare each document of the dataset to its individual set of candidate documents.

Given the size of the result set (approx. 6M document pairs) and our primary goal of searching for undiscovered cases of plagiarism, we employed several filters to focus our manual investigations on the most critical similarities. To eliminate cases, in which authors likely reused own content, we excluded document pairs that shared at least one author. This exclusion prevents the identification of potential self-plagiarism. Similarly, we pruned document pairs, for which the older document cites the newer document, to reduce results in which authors reproduced previous work with due attribution. We make these restrictions for two reasons. First, the definition of what constitutes self-plagiarism varies greatly in different research fields and even for different venues. The vagueness of the problem definition prevents a well-founded assessment of the retrieved documents. Second, because we analyze all documents in the dataset, the number of results is much larger than in the typical PD scenario, i.e., analyzing a single input document.

Since we are particularly interested in the benefit that a math-based similarity assessment can add to a combined approach, we excluded documents with a Histo score below 0.25. i.e., with little math-based similarity, and sorted the remaining results according to the GIT score in descending order. To not exclude cases, in which documents contained unequal amounts of identifiers, e.g., because one document is significantly shorter (cf. \Cref{sec.detailed_analysis}), we did not require a Histo score above the significance threshold of 0.56 but only a score that is greater or equal to 0.25. 

\begin{table}[t!]
\centering
\caption{Top-ranked documents in exploratory study.}%
\label{tab.ratingCases}%
\setlength{\tabcolsep}{2pt} %
\begin{tabular}{l|cccccccccc}
\toprule
\textbf{Rank} & 1 & 2 & 3 & 4 & 5 & 6 & 7 & 8 & 9 & 10 \\ \midrule
\textbf{Case} & C3 & C11 & C12 & C13 & C10 & C14 & C15 & C16 & C17 & C18 \\
\textbf{Rating} & Plag. & Susp. & CR & FP & Plag. & FP & CR & CR & CR & CR \\
\end{tabular}
\vspace{-0.5cm}
\end{table}

\Cref{tab.ratingCases} shows the ten top-ranked document pairs and our rating of the observed similarities. We use the abbreviated ratings 'Plag.' for confirmed cases of plagiarism, 'Susp.' for suspicious content similarity, 'CR' for notable but legitimate content reuse and 'FP' for false positives, i.e., documents with insignificant content overlap.

The highest ranked document pair is the confirmed case of plagiarism C3. The author of the retracted paper copied three geometric proofs with few changes from a significantly longer paper, thus resulting in a high GIT (0.78) but low Histo score (0.29). Another confirmed case of AP (C10) was retrieved at rank 5. The main contribution of the retracted paper in C10, a model in Nuclear Physics, was taken from the source paper while partially renaming identifiers. Almost the entire mathematical content of the retracted paper overlaps with the source document, resulting in the highest Histo score (0.85) in our exploratory study. The differences of identifiers in the source document and the retracted document result in a lower but still suspiciously high GIT score (0.63).

The later document in C11 (rank 2) is a mixture of idea reuse and content reuse. The author of the later paper reused the argumentative structure, sequence of formulae, several of the cited sources, many descriptions of formulae, and non-trivial remarks about the implications of the research from the earlier paper. By doing so, the author of the later paper derived a minor generalization of an entropy model for a specific type of black holes introduced in the earlier paper. The later paper cites other papers by the author of the earlier paper but not the earlier paper itself. We contacted the author of the earlier paper about our findings. In his view, the later paper \textit{"certainly constitutes a case of plagiarism"}. In coordination with the author of the earlier paper, we contacted the journal that published the later paper. The journal's editorial board currently examines the case. Since the journal has not published an official determination about the legitimacy of the paper, we classify the document as suspicious. This case exemplifies the benefits of a combined math-based, citation-based, and text-based similarity analysis. Only a combined analysis reveals the full extent of content similarity that encompasses approx. 80\% of the paper's content.

The five cases of legitimate content reuse (C12, C15, C16, C17, and C18) exhibit similar characteristics. In all five cases, the authors of the later papers reproduce and properly cite extensive mathematical models proposed in the earlier papers. HyPlag failed to recognize the citations and to exclude the document pairs due to two challenges. First, the use of severely abridged citation styles, e.g., only stating the author name(s) and the arXiv identifier of a paper. Second, some authors cite the arXiv preprint of a paper, whereas other authors cite the journal version. The journal versions regularly exhibit differences in the order of authors and the title compared to the respective arXiv preprints. Both cases were not handled correctly by our preprocessing pipeline (cf. \Cref{sec.mathpd.pre}). Clearly, we need to improve our procedures for extracting and disambiguating such challenging references in STEM documents.

However, retrieving these five cases at top ranks is justified given the overlap in mathematical content (typically multiple pages). We expect that reviewers would like to be made aware of such content overlap, e.g., to verify the correct citation of the previous work.

The two false positives, C13 (rank 4) and C14 (rank 6) that \mbox{HyPlag} retrieved reveal potential improvements for the math-based similarity measures. C13 comprises of two papers in Combinatorics that contain long lists of all possible combinations of the identifiers $a$, $b$, and $c$ according to a set of production rules. Similarly, C14 comprises two documents that analyze partition functions and contain long matches entirely made up of the identifiers $p$ and $q$ that occur in large quantities within unrelated formulae. 

To increase the effectiveness of the math-based similarity measures and prevent such false positives, we plan to devise measures that are confined to individual formulae. Likewise, we plan to research how an assessment of structural and semantic similarity of formulae can be adapted for the plagiarism detection use case. Research on formula search and other mathematical information retrieval tasks has provided approaches that could prove valuable for the PD scenario \cite{Guidi2016, Meuschke2017b}. The math-based approach to PD is at an early stage of development. Like the early approaches for text reuse detection \cite{alzahrani2012, Meuschke2013}, we investigated basic, computationally efficient feature analysis methods to identify the reuse of identical and slightly different mathematical content. The results of our investigations show that the math-based analysis approaches increase the detection capabilities for STEM documents, particularly when being combined with other similarity assessments.
\vspace{-1em}
\vspace{+0.5cm}
\section{Conclusion \& Future Work} \label{sec.concl}
By reviewing prior research, we showed that semantic, syntactic and cross-lingual PD approaches achieve high detection effectiveness, even for concealed forms of AP. However, these approaches require a high computational effort. The recall level of efficient text-based candidate retrieval methods stagnates. Approaches that analyze nontextual content features in academic documents, such as citations and mathematical content, show promise for being employed as computationally modest methods to retrieve candidate documents and as more elaborate detailed analysis methods. 

The paper at hand extends a pilot study \cite{Meuschke2017b}, in which we explored the potential of analyzing the similarity of mathematical content in the detailed analysis stage of the external PD process. In the current paper, we additionally devised a computationally efficient candidate retrieval stage that analyzes mathematical content features, academic citations, and textual features using production-ready information retrieval technology. Moreover, we created the GIT and LCIS measures, which consider the order of mathematical identifiers, for the detailed analysis stage. We implemented the newly developed math-based measures, as well as established citation-based and text-based measures in HyPlag - a working prototype of a hybrid plagiarism detection system. 

Using HyPlag, we compared the effectiveness of the math-based, citation-based, and text-based PD approaches using confirmed cases of AP. We showed that a simple unification of the modestly sized sets of candidate documents retrieved by each retrieval heuristic achieved perfect recall for the candidate retrieval stage. For the detailed analysis stage, the newly developed GIT measure exceeded the effectiveness of the best performing approach (Histo) in our pilot study and achieved the same effectiveness as the text-based similarity measure in our current study. 

Errors in the acquisition of in-text citations and bibliographic references decreased the effectiveness of the citation-based similarity measures in our experiments. Despite these limitations, citation-based measures added a significant benefit to the hybrid approach, particularly for the candidate retrieval stage. LCCS also performed decently for the detailed analysis stage (MRR=0.60 for our test cases). The error-corrected similarity scores showed that the true effectiveness of the citation-based measures is much higher.

Overall, the combined analysis of math-based and a citation-based similarity identified all cases that a text-based analysis flagged as strongly suspicious. Moreover, the two nontextual detection approaches provided valuable indicators for suspicious document similarity for cases with a comparably low textual similarity. This result indicates that the best detection effectiveness can be achieved by combining the heterogeneous similarity assessments.

In an exploratory study, we showed the effectiveness of analyzing math-based and citation-based similarity for discovering unknown cases of potential AP. We used the GIT and Histo measures in combination with the citation relations of documents to reduce a result set of approx. 6M document pairs to 10 document pairs that we investigated manually. The highest ranked document pair was a confirmed case of AP. The document retrieved at the second rank was rated as an undiscovered case of AP by the author of the apparent source document. The remaining 8 cases include one confirmed case of AP, 5 documents with high but legitimate overlap in mathematical content and 2 false positives. The citation-based filter would have eliminated 5 cases of legitimate content reuse if the bibliographic data had been extracted correctly. These results show the large potential of analyzing mathematical content and academic citations as a complement to text-based PD approaches. 

Our future work must improve the extraction of citation data to leverage the full potential of the citation-based detection approach. We will increase the number of test cases and their degree of obfuscation to further support our results. For this purpose, we are collaborating with a major mathematical publishing service \cite{Schubotz2019}. 

We will also research improvements to the math-based similarity measures. We expect that the performance of math-based heuristics for candidate retrieval can be improved by incorporating positional information about mathematical features. To improve the math-based measures employed for the detailed analysis stage, we will investigate approaches that consider the structural similarity and extend our research on the semantic similarity of formulae \cite{disSigir16}. 

Another promising direction for future research we plan to pursue is the application of \mbox{machine} learning to balance the weights of the similarity measures. Such an approach could train a detection system tailored to the domain-specific properties of a collection.

In summary, we see the integrated analysis of textual and non-textual features as the most promising approach to deter and to detect academic plagiarism in STEM research publications.

\bibliographystyle{ACM-Reference-Format}
\bibliography{_bibliography}

\end{document}